\documentclass[10pt]{article}

\usepackage{amsmath, amssymb, graphicx, float, listings, geometry, fancyvrb}
\usepackage{lmodern}
\usepackage{setspace}
\usepackage{titling}
\usepackage{enumitem}
\usepackage{booktabs}
\usepackage{appendix}

\usepackage{listings}
\usepackage[T1]{fontenc} 

\usepackage{xcolor}  
\usepackage[
    colorlinks=true,
    linkcolor=blue!60!black,      
    citecolor=green!50!black,     
    filecolor=magenta!70!black,   
    urlcolor=orange!80!black,     
    bookmarks=true,
    bookmarksopen=true,
    bookmarksdepth=3
]{hyperref}

\usepackage{siunitx}
\usepackage{tocloft}
\usepackage{longtable}

\newcommand{\sym}

\pretitle{%
  \begin{center}
  \rule{\linewidth}{1.5pt} \\[\bigskipamount]
  \LARGE\bfseries
}
\posttitle{%
  \\[\bigskipamount]
  \rule{\linewidth}{1.5pt}
  \end{center}
}

\preauthor{\begin{center}\large}
\postauthor{\end{center}}
\predate{\begin{center}}
\postdate{\end{center}}

\title{
\vspace{0.5em}
\large
{\scalebox{1.2}{S}TATISTICALLY
\scalebox{1.2}{S}IGNIFICANT 
\scalebox{1.2}{L}INEAR 
\scalebox{1.2}{R}EGRESSION 
\scalebox{1.2}{C}OEFFICIENTS 
\scalebox{1.2}{S}OLELY 
\scalebox{1.2}{D}RIVEN 
\scalebox{1.2}{B}Y 
\scalebox{1.2}{O}UTLIERS 
\scalebox{1.2}{I}N 
\scalebox{1.2}{F}INITE\scalebox{1.2}{-}SAMPLE 
\scalebox{1.2}{I}NFERENCE}
}

\author{
\large Felix Reichel\thanks{\href{mailto:felix.reichel@jku.at}{felix.reichel@jku.at}\\
B.S., Department of Economics, Johannes Kepler University of Linz,
Linz, Austria, 4040 \\
Thanks to T.Cunningham (MSU) this version now fixes a typo in the title.
}}

\date{\large May 2025}

\begin{document}

\maketitle

\begin{center}
\textit{\large \textbf{A Preprint}}
\end{center}

\vspace{1.5em}

\noindent
\begin{abstract}
In this paper, we investigate the impact of outliers on the statistical significance of coefficients in linear regression. We demonstrate, through numerical simulation using \texttt{R}, that a single outlier can cause an otherwise insignificant coefficient to appear statistically significant. We compare this with robust Huber regression, which reduces the effects of outliers. Afterwards, we approximate the influence of a single outlier on estimated regression coefficients and discuss common diagnostic statistics to detect influential observations in regression (e.g., studentized residuals). Furthermore, we relate this issue to the optional normality assumption in simple linear regression \cite{SCHMIDT2018146}, required for exact finite-sample inference but asymptotically justified for large $n$ by the Central Limit Theorem (CLT). We also address the general dangers of relying solely on p-values without performing adequate regression diagnostics. Finally, we provide a brief overview of regression methods and discuss how they relate to the assumptions of the Gauss-Markov theorem.
\end{abstract}

\vspace{1em}

\vspace{1em}


\vspace{0.5em}


\vspace{3em}

\noindent
\textbf{License:} This work is licensed under a 
\href{https://creativecommons.org/licenses/by-sa/4.0/}{Creative Commons Attribution-ShareAlike 4.0 International License (CC BY-SA 4.0)}. 
You are free to share and adapt the material, provided appropriate credit is given and any derivatives are distributed under the same license.

\vspace{2em}

\noindent
\includegraphics[width=0.25\textwidth]{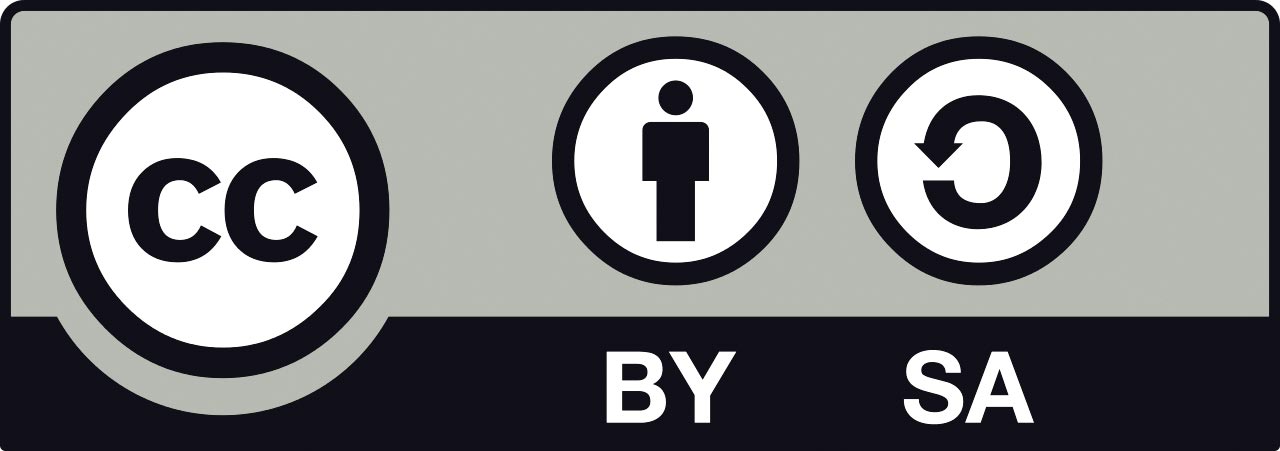}

\newpage
\tableofcontents

\newpage

\section{Introduction}

\noindent Linear regression is a foundational method widely used for modeling due to its simplicity, interpretability, and strong theoretical underpinnings. The classical simple linear regression (SLR) model in  scalar notation is given by:

\begin{equation}
Y_i = \beta_0 + \beta_1 X_i + \varepsilon_i, \quad \varepsilon_i \sim N(0, \sigma^2),
\end{equation}

\noindent where \( \beta_0 \) (the intercept) and \( \beta_1 \) (the slope) are the coefficients capturing the linear relationship between the predictor \( X_i \) and the dependent/outcome/response variable \( Y_i \), for \( i = 1, \dots, n \). Estimation via OLS yields interpretable, closed-form solutions and enables simple statistical inference using t-tests and confidence intervals, assuming that the classical linear model conditions hold. As shown in \cite{wooldridge2019introductory}, under assumptions \textbf{SLR.1} through \textbf{SLR.4} for simple linear regression, the OLS estimates remain unbiased; \textbf{SLR.5}, which assumes normally distributed errors, is only required for valid finite-sample inference using the t-distribution, but can in general be relaxed in large samples due to the Central Limit Theorem. (CLT)

\noindent One of the major advantages of linear regression is the interpretability of the estimated coefficient \( \hat{\beta}_1 \), representing the expected change in the response \( Y \) for a one-unit (assuming no re-scaling e.g. standard-units) increase in the predictor \( X \). Additionally, hypothesis testing on coefficients is conducted using Student’s t-statistic:

\begin{equation}
t = \frac{\hat{\beta}_1}{SE(\hat{\beta}_1)} \sim t_{n - 2}, \quad H_0: \beta_1 = 0,
\end{equation}

\noindent which provides measures of statistical significance with the associated p-values, assuming the normality of the error term or a (sufficiently) large sample size.

\noindent Despite its upsides, a critical limitation of OLS is its strong sensitivity to outliers. The use of the squared loss function \( \sum_{i=1}^n (Y_i - \hat{Y}_i)^2 \) for model fittings causes observations with large residuals to have a disproportionately high influence on the estimated coefficients. Although the sensitivity can in general be informative, it can also lead to highly misleading subsequent inferences and effect calculations. Even a single outlier can distort the estimated coefficient \( \hat{\beta}_1 \), resulting in underestimated standard errors and thus false statistical significance of the predictor variable \cite{belsley1980regression}.

\noindent This paper investigates how outliers can lead to misleading conclusions in regression analysis. Through simulation using \texttt{R}, we demonstrate the fragility of OLS-based inference when a finite sample inference is manipulated by insertion of a single outlier. To mitigate this issue, we additionally fit a robust Huber regression \cite{huber1964robust}.

\noindent In doing so, we aim to demonstrate the limitations of classical OLS regression based on the quadratic loss function for statistical inference and advocate the use of additional robust alternatives and model diagnostic tools.

These include regression diagnostics (for example, residual analysis, leverage, studentized residuals, Cook’s distance \cite{cook1977influential}), model fit statistics (for example, \( R^2 \), F tests), and formal outlier detection methods such as single outlier statistical tests \cite{barnett1978outliers, rstudentbonferroni, Ugah2021} using residuals, to ensure reliable and transparent communication of statistically significant results.

\section{Linear Regression and Statistical Inference}

\noindent We begin by recalling the standard form of the linear regression model in matrix notation:
\begin{equation}
Y = X\beta + \varepsilon, \quad \varepsilon \sim \mathcal{N}(0, \sigma^2 I),
\end{equation}
where \( Y \in \mathbb{R}^n \) is the dependent/outcome/response variable, \( X \in \mathbb{R}^{n \times p} \) is the covariate matrix of predictors (assumed to have full column rank), \( \beta \in \mathbb{R}^p \) is the vector of coefficients to be estimated, and \( \varepsilon \in \mathbb{R}^n \) is its error term, assumed to be independent and identically distributed with zero mean and constant variance \( \sigma^2 \) \cite[Ch. 2]{wooldridge2019introductory}. These assumptions are known as the classical linear model assumptions denoted as \textbf{MLR.1}–\textbf{MLR.5} in \cite{wooldridge2019introductory} in the context of multiple regression for example.

\subsection{OLS Estimation}

\noindent The coefficients \( \beta \) are estimated using OLS, which minimizes the sum of squared residuals:
\begin{equation}
\hat{\beta} = \arg\min_\beta \|Y - X\beta\|^2.
\end{equation}
The solution to this optimization problem is given by:
\begin{equation}
\hat{\beta} = (X^\top X)^{-1}X^\top Y.
\end{equation}
This formula is valid under assumption \textbf{MLR.3} (no perfect multicollinearity), which ensures that \( X^\top X \) is a invertible matrix product \cite[Ch. 3]{wooldridge2019introductory}.

\subsection{Distribution of the OLS Estimator}

\noindent Under optional assumption \textbf{MLR.5}, which states that the error term \( \varepsilon \) is normally distributed, the OLS estimator \( \hat{\beta} \) also follows a multivariate normal distribution:
\begin{equation}
\hat{\beta} \sim \mathcal{N}\left(\beta, \sigma^2 (X^\top X)^{-1}\right).
\end{equation}
This result forms the basis for inference procedures such as hypothesis testing and confidence interval construction \cite[Ch. 4]{wooldridge2019introductory}.

\subsection{Estimating Variance}

\noindent Because \( \sigma^2 \) is unknown in practice, it is estimated using the residuals from the fitted model. Under assumption \textbf{MLR.4} (homoskedasticity), an unbiased estimator of the variance is given by:
\begin{equation}
\hat{\sigma}^2 = \frac{1}{n - p} \|Y - X\hat{\beta}\|^2 = \frac{1}{n - p} \sum_{i=1}^n \hat{\varepsilon}_i^2,
\end{equation}
where \( \hat{\varepsilon} = Y - X\hat{\beta} \) is the vector of residuals \cite[Ch. 3]{wooldridge2019introductory}.

\subsection{Standard Error and Hypothesis Testing}

\noindent The standard error of the estimated coefficient \( \hat{\beta}_j \) is computed as:
\begin{equation}
SE(\hat{\beta}_j) = \sqrt{\hat{\sigma}^2 \left[(X^\top X)^{-1}\right]_{jj}}.
\end{equation}
To test the null hypothesis \( H_0: \beta_j = 0 \), we compute the corresponding \( t \)-statistic as:
\begin{equation}
t_j = \frac{\hat{\beta}_j}{SE(\hat{\beta}_j)}.
\end{equation}
Under the assumptions of the classical linear model—and particularly \textbf{MLR.5}—this statistic follows a Student’s \( t \)-distribution with \( n - p \) or written as \( n - p -1\) (depending on intercept inclusion) degrees of freedom:
\begin{equation}
t_j \sim t_{n - p}.
\end{equation}
The two-sided p-value can be calculated as:
\begin{equation}
\text{p-val.} = 2 \cdot P(T > |t_j|), \quad T \sim t_{n - p},
\end{equation}
which tests whether the coefficient \( \beta_j \) is statistically different from zero in the population regression model \cite[Ch. 4]{wooldridge2019introductory}.

\section{The Effect of a Single Outlier on Regression Coefficients}

OLS is sensitive to outliers. One unusual point can strongly affect the estimated slope, especially if the point has high leverage—that means, if it lies far from the center of the predictor distribution. This can make a non-significant result appear statistically significant \cite{belsley1980regression,cook1977influential}.

\noindent We can show this with a simple example. We generate 100 observations $(x,y)$ in \texttt{R} using the seed: {123} with no true relationship. Model 1, based on this clean data, shows no significant coefficient. In Model 2, we add one extreme outlier. This one point changes the slope to 1.62 and creates a highly significant result. Model 3 uses robust regression (Huber’s M-estimator), which reduces the outlier’s impact and gives a smaller slope again (see Table~\ref{tab:regression_outlier}).

\begin{table}[htbp!]
\centering
\caption{Impact of a Single Outlier on the Statistical Significance of Linear Regression Estimates}
\label{tab:regression_outlier}
\begin{tabular}{
  l
  S[table-format=1.3] S[table-format=1.3]
  S[table-format=1.3] S[table-format=1.3]
  S[table-format=1.3] S[table-format=1.3]
}
\toprule
 & \multicolumn{2}{c}{\textbf{No Outlier}} 
 & \multicolumn{2}{c}{\textbf{With Outlier}} 
 & \multicolumn{2}{c}{\textbf{Robust Regression}} \\
\cmidrule(r){2-3} \cmidrule(r){4-5} \cmidrule(r){6-7}
\textbf{Variable} & \multicolumn{1}{c}{Estimate} & \multicolumn{1}{c}{SE} 
                 & \multicolumn{1}{c}{Estimate} & \multicolumn{1}{c}{SE} 
                 & \multicolumn{1}{c}{Estimate} & \multicolumn{1}{c}{SE} \\
\midrule
Intercept     
    & -0.103 & 0.098 
    & -0.115 & 0.230 
    & -0.162 & 0.097 \\

Coefficient on $x$ 
    & -0.052 & 0.107 
    & \bfseries 1.620*** & \bfseries 0.171 
    & \bfseries 0.184** & \bfseries 0.072 \\

\midrule
Residual Std. Error 
    & \multicolumn{2}{c}{0.971 (df = 98)} 
    & \multicolumn{2}{c}{2.289 (df = 99)} 
    & \multicolumn{2}{c}{0.966 (df = 99)} \\

$R^2$ / Adj. $R^2$  
    & \multicolumn{2}{c}{0.002 / -0.008}
    & \multicolumn{2}{c}{0.476 / 0.471}
    & \multicolumn{2}{c}{--} \\

F Statistic         
    & \multicolumn{2}{c}{0.241 (1, 98)}
    & \multicolumn{2}{c}{\bfseries 89.99\sym{***} (1, 99)} 
    & \multicolumn{2}{c}{--} \\

Observations        
    & \multicolumn{2}{c}{100}
    & \multicolumn{2}{c}{\textit{101}}
    & \multicolumn{2}{c}{\textit{101}} \\
\bottomrule
\end{tabular}

\vspace{1em}
\begin{minipage}{0.95\textwidth}
\small
\textit{Notes:} Model 1 is estimated on clean data. Model 2 adds one outlier, which changes the slope and makes the result statistically significant. Model 3 uses robust regression (Huber) to reduce the influence of the outlier. Robust models often omit $R^2$ or F-statistics because they do not apply directly. See Appendix A for Residual Plots.\\
\textit{Significance levels:} \sym{* }\,$p<0.1$; \sym{** }\,$p<0.05$; \sym{*** }\,$p<0.01$
\end{minipage}
\end{table}

\noindent We can also illustrate the effect mathematically. When a single value in the data changes, the approximate change (using the result of a first-order Taylor approximation) in the OLS estimate is approximately given by:

\begin{equation}
\Delta \hat{\beta} \approx (X^\top X)^{-1} x_i^\top \Delta y_i,
\end{equation}

\noindent where \( x_i \) is the row vector for the \( i \)-th observation. This shows that the change in the estimated slope depends on both the residual size (\( \Delta y_i \)) and the leverage of the (outlier) point. Leverage is defined as:

\begin{equation}
h_i = x_i (X^\top X)^{-1} x_i^\top,
\end{equation}

\noindent where \( h_i \) measures how far \( x_i \) is from the center of the predictor space. High-leverage points can have a disproportionate effect on the fit \cite{belsley1980regression}.

\noindent Robust methods help reduce this sensitivity. Huber’s M-estimator \cite{huber1964robust} uses a different loss function that grows quadratically near zero but linearly in the tails of the distribution, limiting the influence of large residuals. Another robust approach is Least Trimmed Squares (LTS), which fits the model using only the subset of observations with the smallest residuals \cite{rousseeuw1987robust}.

\noindent Robust regression methods provide more stable and reliable estimates when datasets contain outliers.

\section{Single Outlier Tests for Linear Regression Models}

Outlier detection in linear regression models typically revolves around residuals, i.e., differences between observed and fitted values. Several test statistics target unusually large residuals to identify potential outliers.

\subsection{Internally Studentized Residuals}

The internally studentized residual for observation \( i \) is:
\begin{equation}
R_i := \frac{e_i}{\hat{\sigma} \sqrt{1 - h_{ii}}}, \quad \text{where} \quad e_i := y_i - \hat{y}_i,
\end{equation}
\begin{equation}
h_{ii} := \mathbf{x}_i^\top (X^\top X)^{-1} \mathbf{x}_i, \quad \hat{\sigma}^2 := \frac{1}{n - p} \sum_{j=1}^n e_j^2.
\end{equation}
Here, \( h_{ii} \) is the leverage (diagonal of the so-called hat matrix \( H := X(X^\top X)^{-1}X^\top \)). The denominator rescales residuals the by the local variance. 

\noindent Under the assumption of normally distributed errors (\textbf{MLR.5}), \( R_i \) approximately follows a standard normal distribution for large \( n \), but does not exactly follow a \( t \)-distribution. The \textbf{externally studentized residual}, which removes the \( i \)th observation when estimating variance, follows a \( t_{n - p - 1} \) distribution.

\subsection{Maximum Absolute Internally Studentized Residual}

The test statistic for detecting a single outlier is the maximum  studentized residual:
\begin{equation}
R_n := \max_{i = 1, \dots, n} |R_i|.
\end{equation}

Under the null hypothesis \( H_0 \) (no outliers), an approximate critical value can be obtained using the Bonferroni correction. To control the family-wise error rate at level \( \alpha \), we compare \( R_n \) to the quantile of the Student's t-distribution with \( n - p - 1 \) degrees of freedom:
\begin{equation}
\mathbb{P}\left(R_n > t_{1 - \alpha / (2n), n - p - 1}\right) \leq \alpha.
\end{equation}

\subsection{Normalized Maximum Ordinary Residual}

An alternative approach avoids using the hat matrix \( H \) and instead relies on the unadjusted, or raw, residuals. The corresponding test statistic is the normalized maximum ordinary residual:
\begin{equation}
R_n^* = \sqrt{n} \cdot \frac{\max_i |e_i|}{\| \mathbf{e} \|_2} = \sqrt{n} \cdot \frac{\max_i |y_i - \hat{y}_i|}{\sqrt{\sum_{j=1}^n e_j^2}}.
\end{equation}
This statistic highlights large absolute residuals relative to the overall residual magnitude. Since it does not account for leverage, it can be more sensitive to large deviations in response values. However, this also means it may overlook influential outliers associated with high-leverage points.

\subsection{Distributional Properties and Critical Values}

The exact distributions of \( R_n \) (maximum studentized residual) and \( R_n^* \) (normalized maximum ordinary residual) are not analytically tractable. Therefore, critical values are typically estimated using one or more of the following methods:
\begin{itemize}
    \item Bonferroni-adjusted \( t \)-tests applied to individual residuals,
    \item Monte Carlo or permutation tests under the null hypothesis \( H_0 \),
    \item Conservative bounds derived from F-distributions or inverted Student's \( t \)-distributions.
\end{itemize}

\subsection{Upper Bound of Single Outlier Test Statistics}

Ugah et al.~\cite{Ugah2021} derive an upper bound as identical for both as:
\begin{equation}
R_0^* = \sqrt{\frac{(n - p)F_{\alpha/n,1,n-p-1}}{n - p - 1 + F_{\alpha/n,1,n-p-1}}}.
\end{equation}

\begin{figure}[h!]
    \centering
    \includegraphics[width=1.0\linewidth]{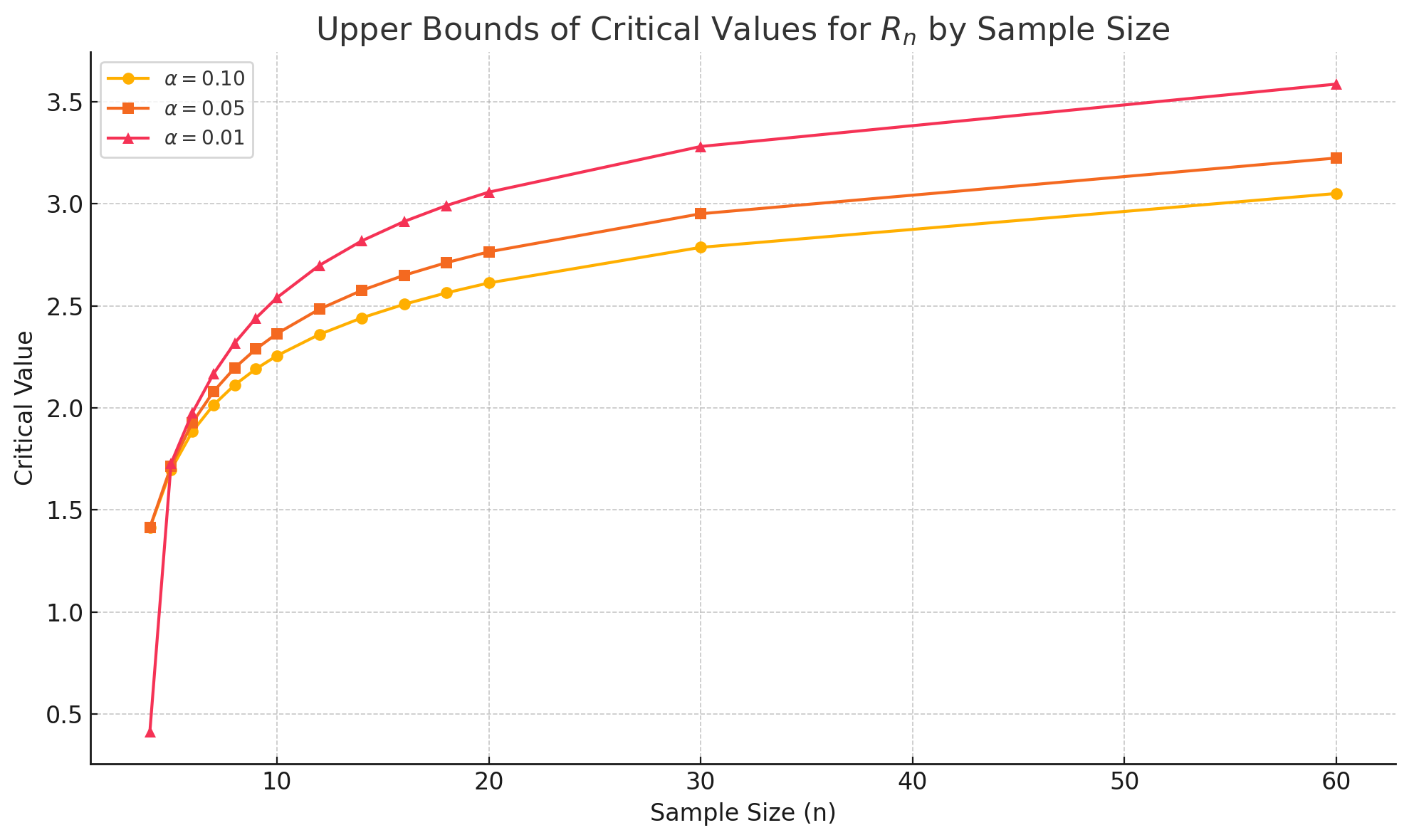}
    \caption{Upper bounds of critical values for \( R_n \) by sample size $n$ and significance level $\alpha$.}
    \label{fig:bounds_plot}
\end{figure}

\clearpage
\newpage

\section{On the Normality Assumption in Regression Analysis}

Recall again \textbf{MLR.5} for exact finite-sample inference. Under this assumption, the least squares estimator
\begin{equation}
\hat{\boldsymbol{\beta}} = (X^\top X)^{-1} X^\top \mathbf{y}
\end{equation}
is unbiased, efficient, and follows:
\begin{equation}
\hat{\boldsymbol{\beta}} \sim \mathcal{N}(\boldsymbol{\beta}, \sigma^2 (X^\top X)^{-1}).
\end{equation}
This enables valid inference using standard test statistics.

\noindent The t-statistic for testing \( H_0: \beta_j = 0 \) is:
\begin{equation}
t_j = \frac{\hat{\beta}_j - \beta_j}{SE(\hat{\beta}_j)} = \frac{\hat{\beta}_j}{\sqrt{\hat{\sigma}^2 (X^\top X)^{-1}_{jj}}} \sim t_{n - p},
\end{equation}
where
\begin{equation}
\hat{\sigma}^2 = \frac{1}{n - p} \sum_{i=1}^n (y_i - \hat{y}_i)^2.
\end{equation}
The residuals \( \hat{\varepsilon}_i \), however, are correlated due to their dependence on the fitted values via the hat matrix \( H = X (X^\top X)^{-1} X^\top \), where \( \hat{\mathbf{y}} = H \mathbf{y} \) and \( \hat{\boldsymbol{\varepsilon}} = (I - H) \mathbf{y} \).

\noindent While OLS remains unbiased and consistent under weaker assumptions (e.g., finite variance and exogeneity), the exact distribution of test statistics like \( t_j \) requires normality.

\subsection*{Outlier Effects on Normality Assumption in Regression Analysis}

Outliers violate this assumption in two key ways:

\begin{enumerate}
    \item \textbf{Non-normality of residuals:} A single outlier induces skewness or kurtosis in the residual distribution, invalidating \( t_j \sim t_{n-p} \). Deviations from linearity are often visible in the Q–Q plot.\\
    $\implies$ Inflated or deflated Type I error rates.
    
    \item \textbf{Distortion of variance estimates:} Outliers distort \( \hat{\sigma}^2 \), affecting directly \( SE(\hat{\beta}_j) \) and thus statistical inference. For high-leverage points \( h_{ii} \) (diagonal entries of the hat matrix), even a small residual \( e_i \) can disproportionately influence:
    \begin{equation}
    SE(\hat{\beta}_j) \propto \sqrt{(X^\top X)^{-1}_{jj}}.
    \end{equation}
    $\implies$ Misestimated p-values, false significance, or masked effects.
    (Robust methods reduce the outlier's influence.)
\end{enumerate}

\noindent Under outlier contamination, the OLS estimator becomes:
\begin{equation}
\hat{\boldsymbol{\beta}} = \hat{\boldsymbol{\beta}}^{(0)} + \Delta(\mathbf{x}_o, y_o),
\end{equation}
where \( \Delta \) increases with leverage \( h_{oo} \) and residual magnitude \( |e_o| \). Thus, \( \hat{\boldsymbol{\beta}} \) can become biased in finite samples and may become inconsistent, particularly if the outlier introduces dependence between the covariates and the errors.

\noindent By the \textbf{Gauss–Markov theorem,} OLS is known as the \textbf{Best Linear Unbiased Estimator (BLUE)} under classical conditions, regardless of normality. In large samples, the \textbf{Central Limit Theorem (CLT)} implies that \( \hat{\boldsymbol{\beta}} \) is approximately normal, allowing for asymptotic inference even when the error distribution is not normal.

\section{An Overview of Regression Methods}

Linear regression relies on the \textbf{Gauss–Markov assumptions}, which ensure that the OLS estimator is the \textbf{BLUE}. These assumptions include linearity of the model, meaning the outcome variable \( \mathbf{y} \) is expressed as a linear combination. The covariate matrix \( X \) must have full column rank so that \( X^\top X \) is invertible, ensuring that the parameter estimates are uniquely defined. Exogeneity is also required, meaning the regressors are uncorrelated with the error term: \( \mathbb{E}[\boldsymbol{\varepsilon} \mid X] = 0 \), which implies \( \operatorname{Cov}(X, \boldsymbol{\varepsilon}) = 0 \). The error terms must be homoscedastic, following a constant variance \( \operatorname{Var}(\boldsymbol{\varepsilon}) = \sigma^2 I \), and they must be uncorrelated across observations, i.e., \( \operatorname{Cov}(\varepsilon_i, \varepsilon_j) = 0 \) for all \( i \neq j \). While these assumptions are sufficient for OLS to be unbiased and efficient, an additional assumption of normally distributed errors, \( \boldsymbol{\varepsilon} \sim \mathcal{N}(0, \sigma^2 I) \), is required for exact finite-sample inference using t-tests and F-tests.

In practice, these conditions are often violated. To address such limitations, various extensions and robust methods have been developed. Below, we outline some of the most widely used regression methods and highlight their relation to the underlying Gauss–Markov assumption.

\subsection{Deming Regression}

Deming regression accounts for measurement error in both \(x\) and \(y\), minimizing orthogonal distances:
\begin{equation}
\min_{\beta_0, \beta_1} \sum_{i=1}^n \frac{(y_i - \beta_0 - \beta_1 x_i)^2}{1 + \lambda}, \quad \lambda = \frac{\sigma_x^2}{\sigma_y^2}
\end{equation}
Useful when both variables are noisy \cite{cornbleet1989deming}.\\
\textbf{Assumptions addressed:} Violates fixed \(x\); assumes homoscedastic, independent errors.

\subsection{Ridge Regression}

Ridge regression applies an \(L_2\) penalty to control variance from multicollinearity:
\begin{equation}
\min_{\beta} \sum_{i=1}^n (y_i - x_i^\top \beta)^2 + \lambda \sum_{j=1}^p \beta_j^2
\end{equation}
\cite{hoerl1970ridge}.\\
\textbf{Assumptions addressed:} Mitigates multicollinearity.

\subsection{Lasso Regression}

Lasso uses an \(L_1\) penalty to induce sparsity:
\begin{equation}
\min_{\beta} \sum_{i=1}^n (y_i - x_i^\top \beta)^2 + \lambda \sum_{j=1}^p |\beta_j|
\end{equation}
Some coefficients may be exactly zero \cite{tibshirani1996regression}.\\
\textbf{Assumptions addressed:} Mitigates multicollinearity.

\subsection{Elastic Net}

Elastic Net combines \(L_1\) and \(L_2\) penalties:
\begin{equation}
\min_{\beta} \sum_{i=1}^n (y_i - x_i^\top \beta)^2 + \lambda_1 \sum |\beta_j| + \lambda_2 \sum \beta_j^2
\end{equation}
Balances sparsity and stability \cite{zou2005regularization}.\\
\textbf{Assumptions addressed:} Mitigates multicollinearity.

\subsection{Robust Regression}

Robust regression reduces sensitivity to outliers by using a different loss function \(\rho(\cdot)\) less sensitive than squared error:
\begin{equation}
\min_{\beta} \sum_{i=1}^n \rho(y_i - x_i^\top \beta)
\end{equation}
Huber’s loss is a common default choice for example \cite{huber1964robust}.\\
\textbf{Assumptions addressed:} Handles non-normality and heteroscedasticity.

\subsection{Quantile Regression}

Quantile regression estimates conditional quantiles by minimizing asymmetric loss:
\begin{equation}
\min_{\beta} \sum_{i=1}^n \rho_\tau(y_i - x_i^\top \beta), \quad \rho_\tau(u) = u(\tau - \mathbf{1}_{\{u < 0\}})
\end{equation}
\cite{koenker1978regression}.\\
\textbf{Assumptions addressed:} Handles heteroscedasticity and non-normality.

\subsection{Principal Component Regression (PCR)}

PCR applies PCA to \(X\), then regresses \(y\) on the top components. It reduces multicollinearity and variance \cite{jolliffe2002principal}.\\
\textbf{Assumptions addressed:} Mitigates multicollinearity.

\subsection{Partial Least Squares (PLS)}

PLS projects \(X\) onto components maximizing covariance with \(y\), often outperforming PCR when predictors are correlated with the outcome \cite{wold1984pls}.\\
\textbf{Assumptions addressed:} Mitigates multicollinearity.

\subsection{LOESS}

LOESS fits local linear models weighted by distance:
\begin{equation}
\min_{\beta_0, \beta_1} \sum_{i=1}^n w_i(x) (y_i - \beta_0 - \beta_1 x_i)^2
\end{equation}
Flexible and nonparametric \cite{cleveland1979robust}.\\
\textbf{Assumptions addressed:} Relaxes linearity; locally handles heteroscedasticity.

\subsection{Spline Regression}

Spline regression uses piecewise polynomials with a smoothness penalty:
\begin{equation}
\min \sum (y_i - f(x_i))^2 + \lambda \int (f''(x))^2 dx
\end{equation}
The smoothing parameter \(\lambda\) controls complexity \cite{de1978practical}.\\
\textbf{Assumptions addressed:} Relaxes linearity.

\subsection{Generalized Linear Models (GLMs)}

GLMs generalize linear models via a link function:
\begin{equation}
g(\mathbb{E}[y_i]) = x_i^\top \beta
\end{equation}
Includes logistic and Poisson models \cite{mccullagh1989generalized}.\\
\textbf{Assumptions addressed:} Handles non-normality and heteroscedasticity.

\subsection{Generalized Additive Models (GAMs)}

GAMs allow additive nonlinear effects:
\begin{equation}
\mathbb{E}[y_i] = \alpha + f_1(x_{i1}) + \dots + f_p(x_{ip})
\end{equation}
Each \(f_j\) is estimated nonparametrically (e.g., splines) \cite{hastie1990generalized}.\\
\textbf{Assumptions addressed:} Relaxes linearity; handles non-normality and mild heteroscedasticity.

\section{Conclusion}

Outliers can inflate the significance of regression coefficients, leading to incorrect inferences. The normality assumption of the residuals is critical for valid t-tests, and its violation---commonly due to outliers---necessitates careful diagnostic checks or the use of robust methods. It is advised to always conduct residual diagnostics and robust regression techniques to assess and mitigate these issues, even post outlier-removal, especially in finite-sample inference. 

\noindent Also, the use of outlier diagnostics utilizing both residuals and leverage is recommendable, such as standardized residuals, leverage plots, Cook's distance, DFBETAS, and influence plots. Normality of residuals can be assessed using tools like the Q–Q plot, Shapiro–Wilk test, or histogram of residuals.

\newpage



\begin{thebibliography}{99}

\bibitem{cornbleet1989deming}
P.~J. Cornbleet and N.~Gochman.
\newblock Incorrect least-squares regression coefficients in method-comparison analysis.
\newblock {\em Clinical Chemistry}, 35(4):584--585, 1989.

\bibitem{hoerl1970ridge}
Arthur~E. Hoerl and Robert~W. Kennard.
\newblock Ridge regression: Biased estimation for nonorthogonal problems.
\newblock {\em Technometrics}, 12(1):55--67, 1970.

\bibitem{tibshirani1996regression}
Robert Tibshirani.
\newblock Regression shrinkage and selection via the lasso.
\newblock {\em Journal of the Royal Statistical Society: Series B}, 58(1):267--288, 1996.

\bibitem{zou2005regularization}
Hui Zou and Trevor Hastie.
\newblock Regularization and variable selection via the elastic net.
\newblock {\em Journal of the Royal Statistical Society: Series B}, 67(2):301--320, 2005.

\bibitem{huber1964robust}
Peter~J. Huber.
\newblock Robust estimation of a location parameter.
\newblock {\em Annals of Mathematical Statistics}, 35(1):73--101, 1964.

\bibitem{koenker1978regression}
Roger Koenker and Gilbert Bassett.
\newblock Regression quantiles.
\newblock {\em Econometrica}, 46(1):33--50, 1978.

\bibitem{jolliffe2002principal}
Ian~T. Jolliffe.
\newblock {\em Principal Component Analysis}.
\newblock Springer, 2nd edition, 2002.

\bibitem{wold1984pls}
Herman Wold, Axel Ruhe, Svante Wold, and William~J. Dunn~III.
\newblock The collinearity problem in linear regression. the partial least squares (PLS) approach to generalized inverses.
\newblock In W.~J. Krzanowski, editor, {\em Multivariate Analysis}, pages 167--198. Academic Press, 1984.

\bibitem{cleveland1979robust}
William~S. Cleveland.
\newblock Robust locally weighted regression and smoothing scatterplots.
\newblock {\em Journal of the American Statistical Association}, 74(368):829--836, 1979.

\bibitem{de1978practical}
Carl de~Boor.
\newblock {\em A Practical Guide to Splines}.
\newblock Springer, 1978.

\bibitem{mccullagh1989generalized}
Peter McCullagh and John~A. Nelder.
\newblock {\em Generalized Linear Models}.
\newblock Chapman and Hall, 2nd edition, 1989.

\bibitem{hastie1990generalized}
Trevor Hastie and Robert Tibshirani.
\newblock {\em Generalized Additive Models}.
\newblock Chapman and Hall, 1990.

\bibitem{rousseeuw1987robust}
Peter~J. Rousseeuw and Annick~M. Leroy.
\newblock {\em Robust Regression and Outlier Detection}.
\newblock Wiley, 1987.

\bibitem{fox2002companion}
John Fox.
\newblock {\em An R and S-Plus Companion to Applied Regression}.
\newblock Sage Publications, 2002.

\bibitem{SCHMIDT2018146}
Amand~F. Schmidt and Chris Finan.
\newblock Linear regression and the normality assumption.
\newblock {\em Journal of Clinical Epidemiology}, 98:146--151, 2018.

\bibitem{wooldridge2019introductory}
Jeffrey~M. Wooldridge.
\newblock {\em Introductory Econometrics: A Modern Approach}.
\newblock Cengage Learning, 7th edition, 2019.

\bibitem{Ugah2021}
Tobias~E. Ugah, Emmanuel~I. Mba, Micheal~C. Eze, Kingsley~C. Arum, Ifeoma~C. Mba, and Henrietta~E. Oranye.
\newblock On the upper bounds of test statistics for a single outlier test in linear regression models.
\newblock {\em Journal of Applied Mathematics}, 2021:5 pages, 2021.

\bibitem{belsley1980regression}
David~A. Belsley, Edwin Kuh, and Roy~E. Welsch.
\newblock {\em Regression Diagnostics: Identifying Influential Data and Sources of Collinearity}.
\newblock John Wiley \& Sons, 1980.

\bibitem{cook1977influential}
R.~Dennis Cook.
\newblock Detection of influential observations in linear regression.
\newblock {\em Technometrics}, 19(1):15--18, 1977.

\bibitem{barnett1978outliers}
Vic Barnett and Toby Lewis.
\newblock {\em Outliers in Statistical Data}.
\newblock Wiley, 1978.

\bibitem{rstudentbonferroni}
Sanford Weisberg.
\newblock Multiple outlier detection in regression using studentized residuals.
\newblock {\em Technometrics}, 22(2):139--144, 1980.

\end{thebibliography}
\clearpage

\appendix
\newpage
\clearpage

\section{Residual Diagnostics}

This appendix presents residual diagnostic plots for three models: (1) a clean OLS model without any outliers, (2) an OLS model with a single high-leverage outlier, and (3) a robust regression model fit to the contaminated data. These plots provide visual evidence of how outliers affect model assumptions and how robust methods mitigate their influence.

\subsection{Residual Plots: SLR OLS Model (Clean Data/No Outlier)}

\begin{figure}[H]
    \centering
    \includegraphics[width=0.95\textwidth]{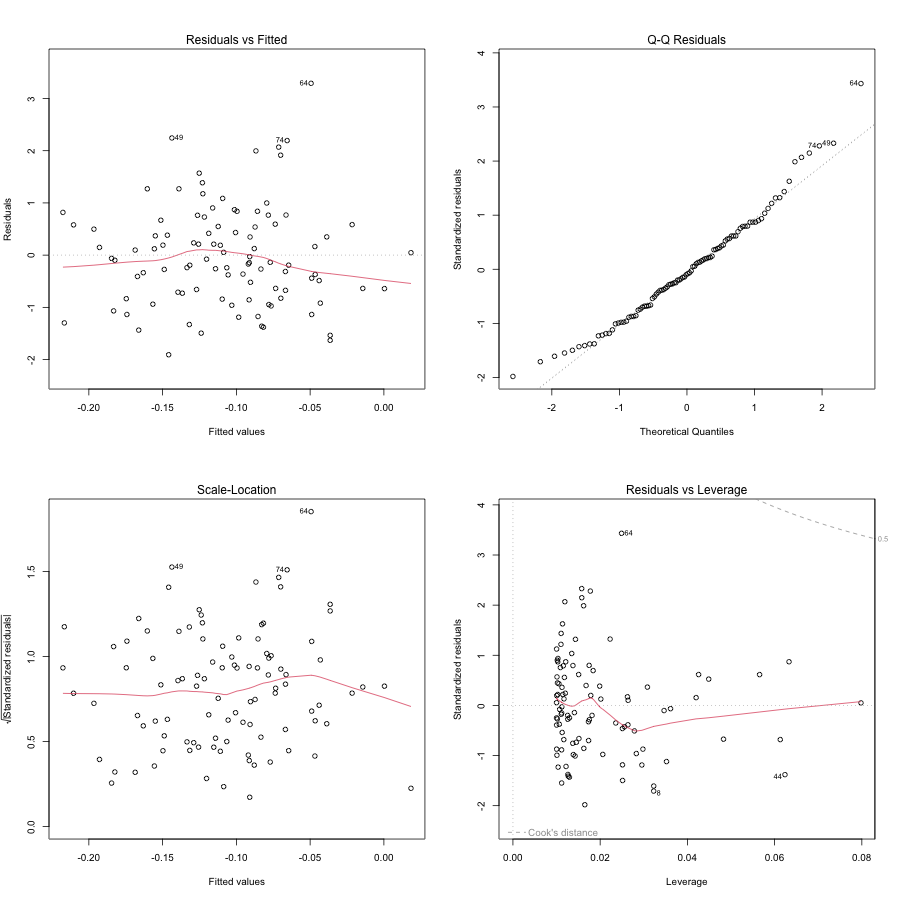}
    \caption{Diagnostic plots for the OLS model. The residuals appear homoscedastic (constant variance), symmetrically distributed, and independent. The Q-Q plot indicates approximate normality, validating the assumptions of OLS.}
\end{figure}

\subsection{Residual Plots: SLR OLS Model With An Outlier}

\begin{figure}[H]
    \centering
    \includegraphics[width=1.0\textwidth]{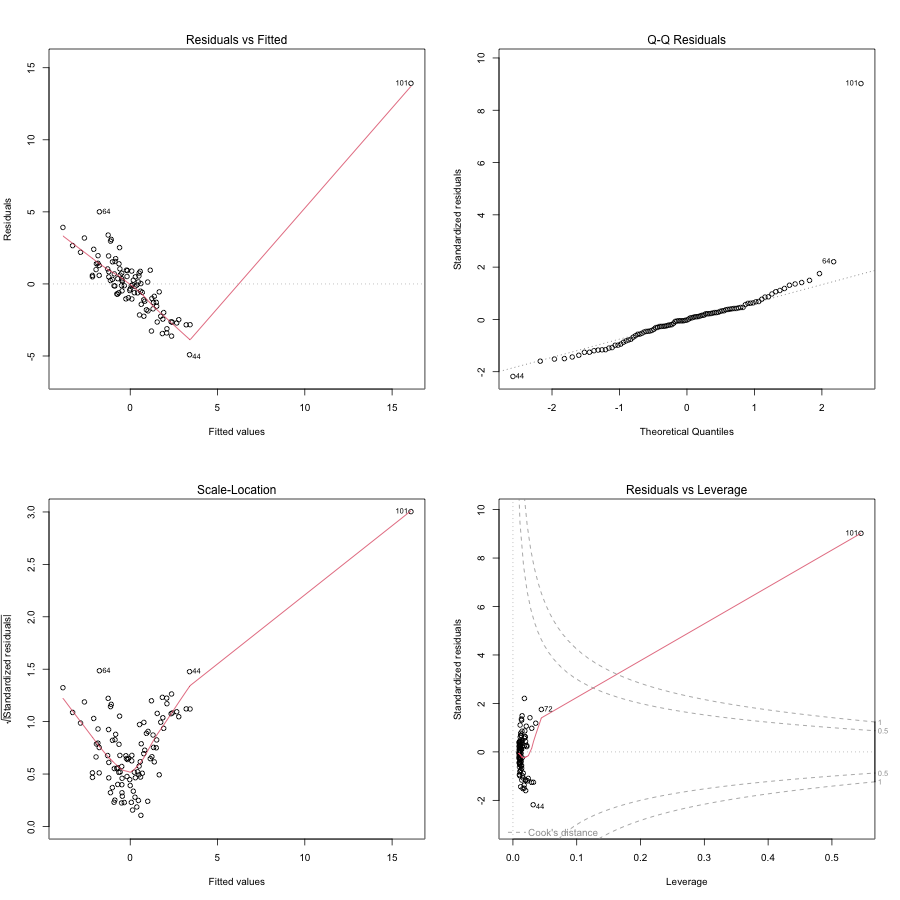}
    \caption{Diagnostic plots for OLS model including a high-leverage outlier. The residuals show clear distortion: heteroscedasticity, skewness, and heavy tails are evident. The Q-Q plot deviates significantly from the normal line, and the residuals vs. fitted plot shows a large residual corresponding to the outlier. This illustrates the breakdown of classical OLS assumptions.}
\end{figure}

\subsection{Residual Plots: SLR Huber Robust With An Outlier}

\begin{figure}[H]
    \centering
    \includegraphics[width=1.0\textwidth]{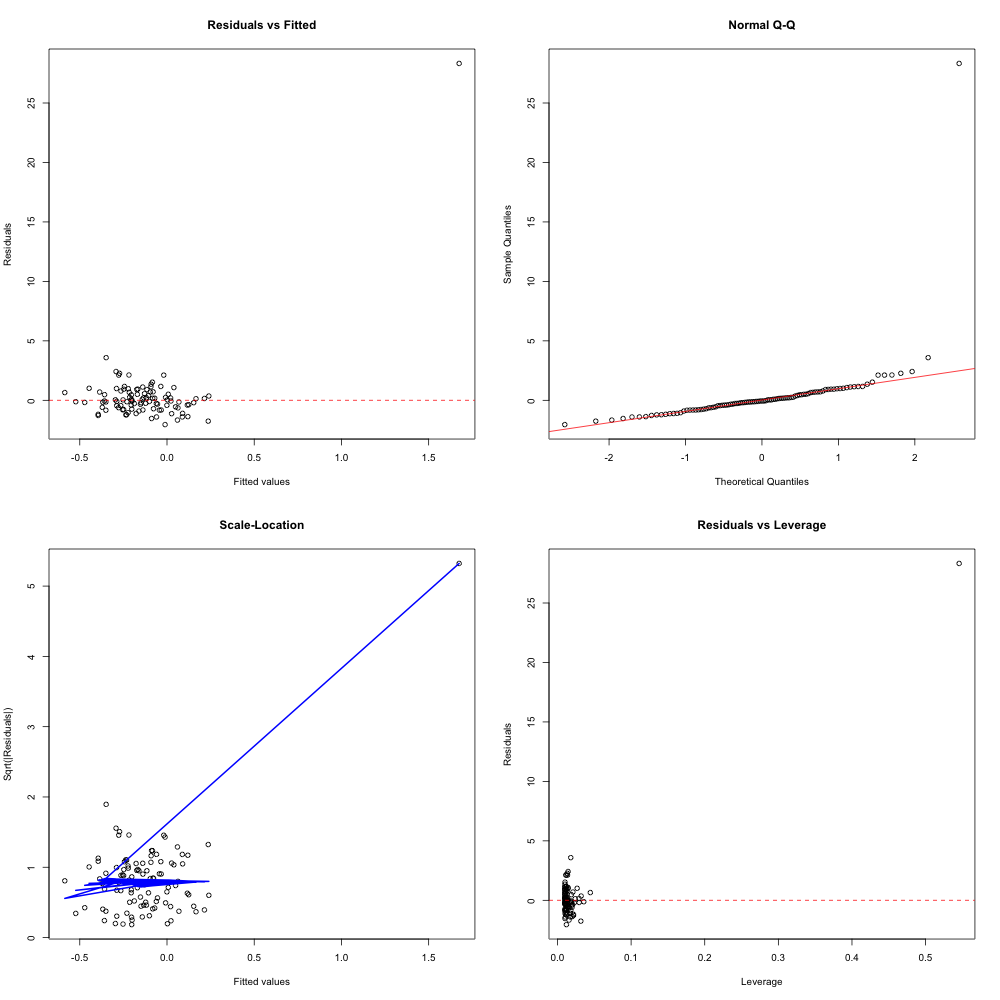}
    \caption{Residuals vs. fitted values for the robust regression model applied to the data with an outlier. Compared to the standard OLS fit (A.2), the residuals are more evenly spread and the extreme influence of the outlier is visibly diminished. This confirms that the robust method effectively downweights the anomalous observation, preserving the integrity of the regression fit.}
\end{figure}

\end{document}